\documentclass[twocolumn,showpacs,preprintnumbers,amsmath,amssymb,pre]{revtex4}
\usepackage{graphicx}
\begin{document}
\title{Segregation in a Fluidized Binary Granular Mixture: Competition between Buoyancy and Geometric Forces}
\author{Leonardo Trujillo$^{1}$, Meheboob Alam$^{2,3}$
 and Hans J. Herrmann$^{1,2}$}                                                                                      
                                                                                      
\affiliation{ $^1$Laboratoire de Physique et M\'ecanique des
Milieux H\'et\'erog\`{e}nes  (UMR CNRS 7636), \'Ecole Sup\'erieure
de Physique et
de Chimie Industrielles, 10 rue Vauquelin, 75231 Paris Cedex 05, France\\
$^2$Institut f\"ur Computeranwendungen 1, Pfaffenwaldring 27,
D-70569 Stuttgart, Germany\\
$^3$Engineering Mechanics Unit, JNCASR, 
Jakkur Campus, Bangalore 560064, India
}
\date{\today}

\begin{abstract}
Starting from the hydrodynamic equations of binary granular mixtures,
we derive an  evolution equation
for the relative velocity of the intruders, which is shown
to be coupled to the inertia of the smaller particles.
The onset of Brazil-nut segregation is explained as a
competition between the buoyancy and geometric forces:
the Archimedean buoyancy force, a buoyancy force due to the
difference between the energies of two granular species,
and two {\it geometric} forces, one {\it compressive}
and the other-one {\it tensile} in nature, due to the {\it size-difference}.
We show that inelastic dissipation strongly affects the phase diagram of the
Brazil nut phenomenon
and our model is able to explain the experimental results of Breu et al.\cite{BEKR03}.
\end{abstract}


\maketitle

\section{Introduction}

Segregation is a process in which a homogeneous mixture of
particles of different species becomes spatially non-uniform by
sorting themselves in terms of their size and/or mass
\cite{HHL98,1,RO,DU,KN,SH98,TH02}.
Monte Carlo simulations of Rosato {\it et al.}
\cite{RO} clearly demonstrated that the larger particles immersed
in a sea of smaller particles rise to the top when subjected to
strong vertical shaking. This is the well-known \emph{Brazil nut
phenomenon} (BNP). It has been explained using the geometrical
ideas of {\it percolation} theory, {\it i.e.} in a vibrated-bed
the smaller particles are more likely to find a void through which
they can percolate down to the bottom, leaving the larger-intruder
at the top\cite{RO,SA89}.
The {\it arching}-effects
\cite{DU}, whereby the larger particle is being supported by the
arches of smaller particles, can help to assist the percolation-driven segregation.
The second mechanism of segregation is the {\it
convective} mean flow in the vibrated bed due to the
formation of convective cells such that the particles move to
the top through the central-axis\cite{KN}.
Recently, another mechanism has been proposed, driven by the {\it inertia}
of the intruder \cite{SH98}, which could explain the {\it
reverse-buoyancy}-effect whereupon a {\it light} but large
particle will sink to the bottom of a deep bed under {\it low-frequency} shaking.
In Ref.\cite{TH02}, a buoyancy-driven segregation mechanism
has been proposed,
drawing a direct analogy with the standard buoyancy forces in a fluid.
For other related issues on segregation, the reader is referred to the
recent review article of Rosato et al.\cite{1}.

The interplay between size and mass has been considered by Hong
\emph{et al.}\cite{HQL01} who found that a downward
intruders' movement occurs as well: \emph{reverse Brazil nut
phenomenon} (RBNP). They proposed a phase diagram for the BNP/RBNP
transition, based on the competition between percolation
and condensation. Recently, Jenkins and Yoon \cite{JY02} investigated
the upward $\Leftrightarrow$ downward transition employing
the hydrodynamic equations for binary mixtures.
The driving mechanism for segregation in the hydrodynamic framework
is presumably different from that of the percolation-condensation idea
and remains unexplained so far.

Employing the Enskog-corrected hydrodynamic equations for binary mixtures\cite{JM},
we investigate the Brazil-nut segregation in a dry fluidized granular mixture
in the absence of bulk convection.
The purpose of this paper is three-fold: firstly, to derive a
time-evolution equation for the relative velocity of a single intruder,
taking into account the non-equipartition of granular energy;
secondly, to explain the driving mechanism for
Brazil-nut segregation in terms of the {\it buoyancy} and {\it geometric} forces.
Lastly, based on a simple model for energy non-equipartition,
we will show how the inelastic dissipation determines
the regimes of BNP and RBNP.

\section{Hydrodynamics of granular mixtures}
The validity of the hydrodynamic approach even in the dense
granular flows has recently been justified via the comparison of theory with various
experiments\cite{LBLG}-- here one has to be careful in choosing the
appropriate constitutive model for pressure, viscosity, dissipation, etc.
The constitutive model that we have used has been validated by performing MD
simulations of binary mixtures\cite{AL}.
We consider a binary mixture of slightly inelastic, smooth
particles (disks/spheres) with radii $r_i$ ($i = l, s$,
where index $l$ stands for large and $s$ for small), mass
$m_i$ and number density  $n_i$. The species mass density is
$\varrho_i({\bf x},t)=m_in_i=\rho_i\phi_i$, where $\rho_i$ is the
material density of species $i$ and $\phi_i$ is its volume
fraction. The total mass-density, $\varrho({\bf x},t)$, and the
total number-density, $n({\bf x},t)$, are just the sums over
their respective species values. The dissipative nature of
particle collisions is taken into account through the normal
coefficient of restitution $e_{ij}$, with $e_{ij}=e_{ji}$ and
$0\le e_{ij}\le 1$.

Assuming unidirectional flow
(${\bf u}_i=(0, v_i(y,t), 0)$; ${\partial}/{\partial x} = 0$, ${\partial}/{\partial z}=0$
and ${\partial}/{\partial y}\neq 0$)
and neglecting viscous stresses, the momentum balance equation
for species $i$ can be written as \cite{JM}
\begin{equation}
\varrho_i\frac{\partial v_i}{\partial t} = -\frac{\partial p_i}{\partial y}
       - \varrho_i {\textrm{g}} + \Gamma_i .
\end{equation}
(Note that the mass balance equations are identically satisfied for
unidirectional flows.)
Here $p_i$ is the {\it partial} pressure of species $i$,
and $\textrm{g}$ the gravitational acceleration acting along the negative $y$-axis;
$\Gamma_i$ is the momentum source term which arises
solely due to the interactions between {\it unlike} particles
and $\sum_{i=l,s} {\Gamma}_i =0$ \cite{JM}.
The assumption of negligible viscous stresses is justified if there is
no overall mean flow in the system, or if the
spatial variation of $v_i(y,t)$ is small.

To obtain constitutive relations for partial pressures
we take into account
the breakdown of equipartition of energy between the two species
(in the equation of state)
as found in many recent theoretical and numerical
studies\cite{JM,BT02,TH02,AL} and also confirmed in
vibrofluidized experiments\cite{FM02}.
We assume that the single particle velocity distribution function
of species $i$ is a Maxwellian
at its own granular energy $T_i$,
where $T_i=\frac{m_i}{d}<{\bf C}_i{\cdot}{\bf C}_i>$,
with $d=2$ and $3$ for disks and spheres, respectively,
${\bf C}_i= {\bf c}_i - {\bf u}$ being the peculiar velocity,
${\bf c}_i$ the instantaneous particle velocity
and ${\bf u}=\varrho^{-1}\sum_{i=l,s}\varrho_i{\bf u}_i$ the mixture velocity.
The equation of state for the partial pressure
of species $i$ can then be written as:
\begin{equation}
   p_i=n_i Z_i T_i, \quad \mbox{with} \quad Z_i= 1 + \sum_{j=l,s} K_{ij}.
\label{eqn_eos}
\end{equation}
Here $Z_i$ is the {\it compressibility} factor of species $i$
and $K_{ij} = \phi_jg_{ij}\left(1+R_{ij}\right)^d/2$, with
$g_{ij}$ being the radial distribution function at contact and
$R_{ij}={r_i}/{r_j}$ the size-ratio. Note that the $K_{ij}$ are
related to the collisional component of the partial pressure,
having a {\it weak} dependence on inelasticity which we neglect, and
$Z_i\to 1$ in the dilute limit $\phi\to 0$. 
We shall return back
to discuss more about the compressibility factor later.

After some algebraic manipulations with the momentum
balance equations and the equation of state, we obtain the following evolution equation
for the relative velocity of the larger particles, $v_l^r=v_l-v_s$,
\begin{eqnarray}
  \varrho_l\frac{\partial v_l^r}{\partial t} &=&
        n_l\left[m_s\left(\frac{Z_l}{Z_s}\frac{T_l}{T_s}\right) - m_l\right] \textrm{g}
        +\left[1+\frac{p_l}{p_s}\right]\Gamma_l 
        \nonumber \\
     & &  + p_l\frac{\partial}{\partial y}\left[\ln\left(
            \frac{p_s}{p_l}\right)\right]
      -\varrho_s\left(\frac{\varrho_l}{\varrho_s}
      -\frac{p_l}{p_s}\right)\frac{\partial v_s}{\partial t}.
\label{eqn_evol1}
\end{eqnarray}
An explicit expression for the momentum source term, ${\Gamma}_l$, can be
obtained using the Maxwellian velocity distribution function \cite{JM}.
\begin{eqnarray}
\Gamma_l &=&
 n_lK_{ls}T\left[\left(\frac{m_s-m_l}{m_{ls}}\right)\frac{\partial}{\partial y}\left(\ln T\right)
 + \frac{\partial}{\partial y}\left[\ln\left(\frac{n_l}{n_s}\right)\right] 
      \right.  \nonumber \\  
  & & \left.  +\frac{4}{r_{ls}}\left(\frac{2m_lm_s}{\pi m_{ls} T}\right)^{1/2}
   \left({v}_s - {v}_l\right)\right],
\label{eqn_msource}
\end{eqnarray}
where $T = n^{-1}\sum_{i=l,s} n_iT_i = \sum_{i=l,s} \xi_i T_i$
is the {\it mixture} granular energy,
$\xi_i=n_i/n$  the number-fraction of species $i$, $m_{ls}=m_l+m_s$ and $r_{ls}=r_l+r_s$.
With additional assumptions of weak gradients in species number densities
and granular energy, and retaining terms of the same order in the
single intruder limit ($n_l << n_s$), the evolution equation can be
considerably simplified to
\begin{eqnarray}
 m_l\frac{{\rm d} v_l^r}{{\rm d} t} &=&
        \left[m_s\left(\frac{Z_l}{Z_s}\frac{T_l}{T_s}\right)- m_l\right] g
     - \frac{4K_{ls}T}{r_{ls}}\left(\frac{2m_lm_s}{\pi m_{ls} T}\right)^{1/2}\!\!\! v_l^r 
       \nonumber \\
       & &   +\left[m_s\left(\frac{Z_l}{Z_s}\frac{T_l}{T_s}\right)-m_l\right]
           \frac{{\rm d} v_s}{{\rm d} t}.
\label{eqn_evol2}
\end{eqnarray}
This is our time-evolution equation for the
relative velocity of a single intruder:
the first term on the right hand side is the net gravitational force
acting on the intruder,
the second term is a `Stokesian-like' drag force and
the third term represents a weighted {\it coupling} with
the inertia of the smaller particles.
It is interesting to recall the work of Shinbrot and Muzzio\cite{SH98}
who argued that the onset of reverse-buoyancy
would crucially depend on the inertia of the smaller particles--
a detailed analysis of eq.~(\ref{eqn_evol2}) with appropriate
boundary conditions is left out for a future investigation.
In this paper, we are only interested in the {\it steady-state}
solution of the above equation.
In typical situations where one can neglect
the last term (e.g. if the intruder is much heavier than the smaller particles),
we end up with the familiar evolution equation where
the inertia of the intruder is being balanced
by the net gravitational force and the drag force.
Only in this case, the interplay between the gravitational
and drag force will eventually decide whether the intruder rises or sinks.
Neglecting transient effects,
the steady relative velocity of the intruder can be obtained from
\begin{equation}
        {v}_l^r =
         \frac{r_{ls}\textrm{g}}{4 K_{ls}}\left(\frac{\pi m_{ls}}{2m_lm_s T}\right)^{1/2}
         \left[m_s\left(\frac{Z_l}{Z_s}\frac{T_l}{T_s}\right)- m_l\right].
\label{eqn_rvelocity}
\end{equation}
Setting this  relative velocity to zero, we obtain the
criterion for the {\it transition} from  BNP to RBNP:
\begin{equation}
   m_s\left(\frac{Z_l}{Z_s}\frac{T_l}{T_s}\right) -m_l = 0,
\label{eqn_scriterion}
\end{equation}
which agrees with the expression  of Jenkins \& Yoon\cite{JY02}
for the case of equal granular energies ($T_l=T_s$).
As such, it is not evident from this expression
{\it what the driving mechanism for segregation is}.
Thus, we need to answer several questions.
Can we recast the segregation criterion in terms
of the well-known Archimedean and thermal buoyancy forces?
Is there any new force, and what could be the
physical origin of such forces?

\section{Driving mechanism: segregation forces}

To understand the {\it origin} of segregation
in the present framework, we now decompose the net gravitational force
in eq.~(\ref{eqn_evol2}) for a {\it single} intruder in the following manner:
\begin{eqnarray}
    F  &=& \textrm{g}
   \left[(\rho_s-\rho_l)V_l + m_s\left(\frac{T_l}{T_s}-1\right)\frac{Z_l}{Z_s}
      \right. \nonumber \\
    & & \left. + m_s\left(1-\frac{V_l}{V_s}\right)
       + m_s\left(\frac{Z_l}{Z_s}-1\right)\right],
\label{eqn_sforce1}
\end{eqnarray}
where $V_i$ is the volume of a particle of species $i$.
The first term, $F_B^A= V_l(\rho_s - \rho_l){\textrm{g}}$,
is the effective Archimedean buoyancy force
which arises due to the weight of the {\it displaced} volume of the intruder ($V_l$).
The second term, $F_B^T\propto (T_l - T_s)$, represents the buoyancy
force due to the difference between the two species granular energies.
This, being an {\it analog} of the thermal buoyancy,
may be termed  the {\it pseudo-thermal} buoyancy force.

There are two more terms in eq.~(\ref{eqn_sforce1}) which do not appear to be
related to standard buoyancy arguments.
The third term  is {\it negative definite}, and vanishes
identically if the intruder and the smaller particles have the {\it same size}.
Note that $\epsilon_v^{st}=(V_l/V_s-1)$ is the {\it volumetric} strain.
Thus, $F_{ge}^{st}=-m_s{\textrm{g}}\epsilon_v^{st}$ is a
{\it static compressive} force to
overcome the  barrier of the {\it compressive} volumetric strain arising
out of the {\it size-disparity} between the intruder and the smaller particles.

The fourth term in eq.~(\ref{eqn_sforce1}), $\propto (Z_l/Z_s-1)$,
vanishes in the dilute limit $\phi\to 0$.
It can be verified that $(Z_l/Z_s- 1)$ also vanishes
identically, irrespective of the total volume fraction,
{\it if the particles are of the same size}.
(Note that we have neglected the weak-dependence of $Z_i$
on inelasticity.)
Thus, the {\it origin} of this force is also tied to the {\it size-disparity}
as in the third term $F_{ge}^{st}$.
An interesting {\it physical} interpretation can be made  if we consider
the dense limit with a single intruder ($\phi_l<<\phi_s$):
$(Z_l/Z_s-1) \propto R_{ls}^d$ for $R_{ls}>>1$.
Hence $\epsilon_v^{dyn}=(Z_l/Z_s-1) \geq 0$ can be associated with
a weighted {\it volumetric} strain, {\it tensile} in nature.
Thus,  $F_{ge}^{dyn}=m_s\textrm{g}\epsilon_v^{dyn}$
is a {\it dynamic} tensile force that arises from
the {\it excess} pressure difference due to the
{\it nonideal} (collisional) interactions
between the intruder and the {\it displaced} smaller particles.

Thus, the {\it geometric} effects due to the {\it size-disparity}
contribute two new types of segregation forces:
\begin{equation}
 F_{ge} = F_{ge}^{st} + F_{ge}^{dyn} =
   - m_s\left(\epsilon_v^{st}-\epsilon_v^{dyn}\right)\textrm{g},
\label{eqn_geoforce}
\end{equation}
the former is a static, compressive force and the latter is a dynamic,
tensile force. On the whole, the collisional interactions
help to reduce the net compressive force that
the intruder has to overcome.

A question naturally arises as to whether we could get back the
standard Archimedes law from eq.~(\ref{eqn_sforce1}) if we take
the corresponding fluid limit, i.e. a large particle being
immersed in a sea of small particles with $r_l >> r_s$. In this
limit it immediately follows
that $F_{ge}^{dyn} \to m_s (V_l/V_s -1) = - F_{ge}^{st}$ and hence
$F_{ge} \equiv 0$. Thus, the net gravitational force on a particle
falling/rising in an otherwise quiscent fluid (at the same
temperature) is nothing but the standard Archimedean buoyancy
force, $F = F_B^A = g(\rho_s - \rho_l)V_l$. It is worth recalling
that when there is {\it no} size-disparity ($r_l=r_s$), the
geometric forces are identically zero. Hence the behaviour of a
heavier particle in a sea of equal-size lighter particles is
similar to that of a particle in a fluid.

\begin{figure}
\includegraphics[width=7.0cm]{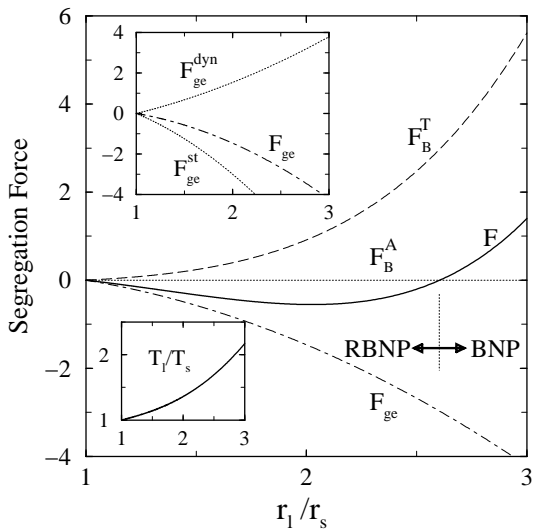}\\
\caption{
Variations of segregation forces ($F/m_s g$) with the size-ratio for
$\rho_l=\rho_s$ at $e=0.9$; see text for other details.
The upper inset shows the corresponding
static and dynamic contributions to the total geometric force
($F_{ge}=F_{ge}^{st} + F_{ge}^{dyn}$).
The lower inset shows the variation of $T_l/T_s$ with the
size-ratio\cite{BT02}.
}
\label{f1}
\end{figure}

To clarify our segregation mechanism, we show the variations of different segregation
forces with the size-ratio in Fig.~\ref{f1}
for the two-dimensional case of {\it equal density} particles ($\rho_l=\rho_s$)
in the single intruder limit ($\phi_l/\phi_s=10^{-8}$)
at a total solid fraction of $\phi=0.7$,
with the restitution coefficient being set to $0.9$.
For illustrative purposes, we have calculated the energy ratio,
$T_l/T_s$, (see the lower inset in Fig.~\ref{f1})
from the model of Barrat and Trizac\cite{BT02}.
For this case, the Archimedean buoyancy force is identically zero,
and the total geometric force remains negative, as seen from the
upper inset in Fig.~\ref{f1}.
The pseudo-thermal buoyancy force is, however, positive.
Thus, the competition between the
pseudo-thermal buoyancy force and the geometric forces
leads to a critical size-ratio above which the intruder will rise for this case.
(For the corresponding purely elastic case ($e=1$ and $F_B^T = 0$),
the net force is $F\equiv F_{ge} <0$ and hence the larger particle will
sink to the bottom.)
This mechanism holds also for the more general case
($\rho_l\neq\rho_s$ and $F_B^A\neq 0$) for which the total buoyancy
force ($F_B=F_B^A+F_B^T$) competes with
the geometric forces ($F_{ge}=F_{ge}^{st}+F_{ge}^{dyn}$)
to determine the transition from BNP to RBNP;
the inclusion of dissipation merely affects the location of the
transition point (see Fig.~\ref{f2} and the discussion below, for details).

\section{Phase diagram and discussion}
\begin{figure}
\includegraphics[width=7.0cm]{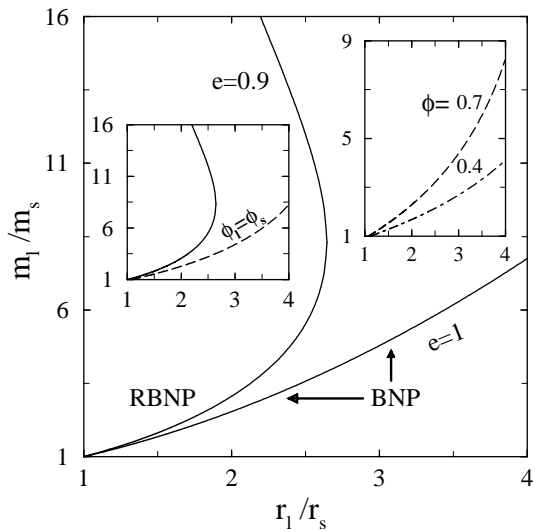}\\
\caption{Phase diagram for BNP/RBNP in two dimensions:
$\phi=0.7$ and $\phi_l/\phi_s=10^{-8}$.
Left inset: phase diagram with $e=0.9$, $\phi_l/\phi_s=10^{-8}$
(solid curve) and  $\phi_l/\phi_s=1$ (dashed curve).
Right inset: phase diagram with $e=0.9$, $\phi_l/\phi_s=1$,
$\phi=0.7$ (dashed curve) and $\phi=0.4$ (dot-dashed curve).
}
\label{f2}
\end{figure}

A typical phase diagram in the  single intruder limit
($\phi_l/\phi_s=10^{-8}$), delineating the regimes between BNP and
RBNP, is shown in Fig.~\ref{f2} for the two-dimensional case, with other
parameters as in Fig.~\ref{f1}. (The qualitative features of the
corresponding phase diagram for the three-dimensional case are
similar.) Focussing  on the purely elastic case ($e=1$), we note
that a transition from BNP to RBNP can occur following two paths
(denoted by two arrows), one along the {\it constant mass-ratio}
with decreasing  size-ratio and the other along the {\it constant
size-ratio} with increasing  mass ratio. In both cases, the
Archimedean buoyancy force balances the net geometric forces at
the transition point.

Comparison between the elastic ($e=1$) and inelastic ($e=0.9$) cases
in Fig.~\ref{f2} clearly shows that the non-equipartition of granular energy,
responsible for the pseudo-thermal buoyancy force $F_B^T$,
has a dramatic effect in reducing the regime of RBNP, and decreasing the
value of $e$ reduces the size of this regime further.
For the case of a mixture with equal volume fractions
($\phi_l=\phi_s$), however, the regime of RBNP
is much larger as seen from the left-inset of Fig.~\ref{f2}.
This observation is in qualitative agreement  with
the recent experimental results of Breu {\it et al.}\cite{BEKR03}
who found that the `{\it reverse Brazil-nut effect is
completely destroyed} if $\phi_s >> \phi_l$'.

The right-inset of Fig.~\ref{f2} shows that the size of the RBNP-regime
also increases with decreasing  overall mean volume fraction.
Since in vibrated-bed experiments increasing the shaking amplitude
is equivalent to decreasing the mean volume fraction, our
observation explains another interesting result of Breu {\it et
al.}\cite{BEKR03} that for a given mixture with specified size-
and mass-ratio, {\it the final state is  that of RBNP at
sufficiently high accelerations} (see Fig.~\ref{f2} in
Ref.\cite{BEKR03}).

We need to point out that calculating the energy ratio
($T_l/T_s$,\cite{BT02}) we made the assumption that $e_{ij}=e$.
Using a variable restitution coefficient, our phase-diagram at
large mass-ratios will be modified, but the proposed segregation
mechanism and the qualitative features of the phase-diagram remain intact. 
Even though the model of Barrat \& Trizac\cite{BT02}
is strictly valid for a homogeneous mixture with stochastic-driving,
it has recently been verified in vibrofluidized-bed experiments
under strong shaking\cite{FM02}.

To compare our segregation mechanism with others, we note that the
scaling of the geometric forces ($\propto R_{ls}^d$) suggests that
they can be compared to the effective percolation force of Rosato
et al.\cite{RO}, and hence we have a competition between buoyancy
and percolation forces.
In the percolation-condensation mechanism
of Hong et al.\cite{HQL01}, the condensation is driven by the
two-species having  different energies.
If one equates their driving force due to condensation-tendency
with an effective buoyancy force, then our mechanism could be equivalent
to that of Hong et al. 
However, there is no direct one-to-one analogy between our
hydrodynamic segregation mechanism and 
the percolation-condensation mechanism.

In conclusion, we have identified four different types of segregation forces:
apart from the Archimedean buoyancy force and an analog of the thermal buoyancy
force, there are two additional forces, the origin
of both is tied to the size-disparity between the intruder
and the smaller particles. We have demonstrated that
the competition between the buoyancy and geometric
forces determines the onset of segregation in the present scenario,
and the inclusion of the pseudo-thermal buoyancy force
(due to inelastic dissipation) further enhances the possibility of BNP.
While the possibility of RBNP is rather limited in the single intruder limit,
even at moderate dissipation-levels,
either increasing the relative volume fraction of the intruders
or decreasing the mean volume fraction
enhances its likeliness as in the experiments of Breu et al.\cite{BEKR03}.

\begin{acknowledgments}
M.A. acknowledges the financial support from the AvH foundation,
and discussions with Stefan Luding on related topics.
\end{acknowledgments}


\end{document}